\documentstyle[epsf]{article}

\newcommand{\be}{\begin{equation}}
\newcommand{\ee}{\end{equation}}

\begin{document}

  
\vspace{5mm}
\begin{center}

{\LARGE \bf Mixing of scalar glueballs and flavour-singlet  scalar
mesons }\\[10mm] 
 {\large\it UKQCD Collaboration}\\[3mm]
 
 {\bf    C. McNeile and   C.~Michael  \\ 
Theoretical Physics Division, Dept. of Mathematical Sciences, 
          University of Liverpool, Liverpool L69 3BX, UK }\\[2mm]

\end{center}

\begin{abstract}
 
We discuss in detail the extraction of  hadronic mixing strengths from
lattice studies. We apply this to the mixing of a scalar glueball and a
scalar meson in the quenched approximation.   We also measure
correlations appropriate for flavour-singlet scalar mesons using 
dynamical quark configurations from UKQCD. This enables us to compare
the results from the quenched study of the mixing with the direct
determination of the mixed spectrum. Improved methods of  evaluating the
disconnected quark diagrams are also presented.

\end{abstract}

\section{Introduction}

 Lattice techniques are well developed to describe mass spectra. What is
 much less well studied are hadronic transitions. Here we shall
concentrate on  purely hadronic transitions such as glueball mixing with
scalar mesons, string breaking, flavour singlet pseudoscalar mass
generation,  etc. In full QCD studies on a lattice, much as in
experiment, one will obtain the  mass values of the resulting mixed
states. By varying quark masses and the  number of quark flavours, one
may be able to go beyond experiment and so help to  substantiate or 
vitiate phenomenological models.  

 Within a quenched or partially quenched  lattice approach, one can in
principle learn much more: obtaining estimates  of the mixing strengths
themselves. This is the approach that we analyse in detail.  We then
apply it to the mixing of a glueball with a scalar meson.  This  is of
considerable phenomenological interest: the fate of the glueball  is
widely debated~\cite{weinmix,close}. 

 As a counterpoint to our quenched study of this  mixing, we also
determine the mixed spectrum directly for two flavours of degenerate sea
quark. This provides a check on our approach  and, incidentally,
indicates evidence for a surprisingly light scalar state  at the lattice
spacing we employ.  

 We include an appendix giving details of the variance reduction technique 
relevant to determining disconnected fermionic loops which are needed in 
our study of flavour singlet mesons.

\section{Lattice Analysis}

 Here we discuss the formalism on a lattice to extract hadronic mixing. 
To set the scene, the variational approach is first summarised and  the
simpler case of weak/electromagnetic matrix elements is reviewed.  Then
we discuss hadronic mixing matrix elements. 

\subsection{Variational methods}

 Consider a hadronic correlator $C_{ij}(t)$ where $t$ is the lattice
separation  in the time direction and $i,\ j$ label the type of operator
used to create/destroy the hadron (eg. whether local, fuzzed, etc). We
assume there are $N$ types of operator.

In an ideal world of infinite statistics   the matrix $C_{ij}(t)$ of
correlations  can be written in terms of the eigenstates of the transfer
matrix as 
 \be 
C(t) = A^{T} e^{ - m t} A
 \ee 
 where the intermediate state sums are over all ``particles'' allowed.

However in practice, we only have a small  matrix $C$ and because of
noise there is  a truncation of the sum in the above equation.

One standard approach to circumvent these problems is the variational
method. This can be motivated by maximising 
 \be
 u_i C_{ij}(t)\ u_j
 \ee 
 subject to constant $ u_i C_{ij}(t-1) u_j$ which leads to the
generalised eigenvalue equation 
 \be
  C_{ij}(t)\ u^\alpha_{j}=\lambda^\alpha C_{ij}(t-1)\ u^\alpha_{j}
  \ee
 with $\alpha = 0,\dots N-1$ and where the ground state with $alpha=0$
has the largest eigenvalue $\lambda^0$. It is usual to relate these
eigenvalues to masses (energies in general)  by $\lambda^\alpha =
e^{-m_{\alpha}}$. Given exact data, $N$ eigenvalues can be determined.  

 In order to isolate a particular state, usually the ground state, one
can use  these variational eigenvectors to form a new basis. This is 
 \be
    \bar{C}_{\alpha \beta}(t)=u^{\alpha}_i C_{ij}(t)u^{\beta}_j
 \ee
 Then at times $t$ and $t-1$, $\bar{C}_{\alpha \beta}$ will be diagonal
and  the diagonal elements will decrease like $\lambda^{\alpha}$ (ie as
$e^{-m_{\alpha}}$) as time increases from $t-1$ to $t$. 

 A typical use of the variational method to extract the ground state
signal is to form this variational basis using a low value of $t$ where
statistical errors are relatively small  and then to explore the $t$
dependence of $\bar{C}_{00}(t)$ at larger $t$- values to  extract the
asymptotic behaviour. Note that the finite ($N$ state) variational basis
derived at finite $t$ will not match exactly to the true spectrum of
excited states and hence $\bar{C}_{00}$ will  have some small remaining
contamination of excited states. Indeed  it is possible to use the
variational estimate of the mass gap to the first excited  state
($m_1-m_0$) to control this extrapolation to large $t$ to determine 
$m_0$.

 In order to make contact with the alternative procedure of fitting $C$ 
directly to $M$ states over some range of $t$:
 \be
  C_{ij}(t) =\sum_{\alpha=0}^{M-1} c^\alpha_i c^\alpha_j e^{-m_\alpha t}
 \ee
 we note that $u$ corresponds to the right eigenvectors of the 
non-symmetric matrix $C^{-1}(t-1)C(t)$ and one can introduce left
eigenvectors $v$ which satisfy
 \be
  v^\alpha_i = C_{ij}(t-1)\  u^\alpha_j
 \ee
 and are orthogonal to the right eigenvectors 
 \be
   v^\alpha_i u^\beta_i=\delta_{\alpha \beta}
 \ee
 In terms of these eigenvectors, we have 
 \be
  C_{ij}(t) =\sum_{\alpha=0}^{N-1} v^\alpha_i v^\alpha_j \lambda^\alpha
 \ee
 and
 \be
  C_{ij}(t-1) =\sum_{\alpha=0}^{N-1} v^\alpha_i v^\alpha_j 
 \ee
  Thus we see that the variational method corresponds to making an exact
 fit to the data at $t$ and $t-1$ with $N$ states with the eigenvalues
giving the masses and the left eigenvectors  $v$ are the couplings.

 \subsection{Matrix elements: Operator insertions}

  Here we consider first the simpler case where a three point function 
is evaluated with an explicit operator (current) at an intermediate time
$t_1$. This is the case of weak and electromagnetic current insertions
and also for  some hadronic studies. One example is the study of
semi-leptonic decays~\cite{IW}. In general the quantum numbers of
the states propagating before and after the insertion may  be different.
 We shall assume that the spectrum of states is discrete.  In a finite
spatial volume,  the two particle spectrum will indeed be discrete and
it is  possible to make use  of this to explore relevant matrix
elements~\cite{ll} for two particle systems. 

  We study the general behaviour of the three point function with
insertion at $t_1$  where $t$ is the lattice separation  in the time
direction and $i,\ j$ label the type of operator used to create/destroy
the hadronic state at times 0 and $t$ where there are $N,\ N'$ types of
operator respectively.  We  include $M$ states with masses(energies)
$m_{\alpha}$ from  time 0 to $t_1$ and  $M'$ states with
masses(energies) $m'_{\beta}$ from $t_1$ to $t$:
 \be
X_{ij}(t_1,t) =\sum_{\alpha=0}^{M-1}\sum_{\beta=0}^{M'-1} c^\alpha_i  
e^{-m_\alpha t_1} x_{\alpha \beta}  e^{-m'_\beta (t-t_1)} d^{\beta}_j
 \ee

 The task is usually to determine the matrix element $x_{00}$
corresponding to  the ground state hadrons.  One can employ a fit to the
above expression along with fitting the two point functions ($C$ with
couplings $c$  and masses $m$ and $C'$ with couplings $d$ and masses
$m'$).  Care should be taken to include sufficient states $M$ and $M'$.  A
sensible criterion is that $M$ should be chosen so that a good fit is
obtained to the  two point correlator over time interval $t_1$ for the
appropriate hadron  with creation/destruction operators as used in the
three point function for that hadron  and   $M'$ should be chosen
likewise so that a good  fit was obtained for the appropriate two point
correlator over time interval $t-t_1$.

Another way to approach this analysis is to  use of the variational
technique discussed above. Then with a variational basis for the
appropriate two point  functions (of size $N$, $N'$ respectively) one
can form 
 \be
    \bar{X}_{\alpha \beta}(t_1,t)=u^{\alpha}_i X_{ij}(t_1,t)u'^{\beta}_j
 \ee

 This expression will have non-zero off-diagonal elements in general
since  the operator insertion need not be diagonal in the spectrum
basis.  In  this variational approach, one can extract $x_{00}$ by
taking a ratio to the variational ground state two point functions
determined as above
 \be
  x_{00}= \bar{X}_{00}(t_1,t)/\sqrt{\bar{C}_{00}(t) \bar{C}'_{00}(t)}
 \ee
 where this ratio should be independent of $t_1$ provided $t_1$ and $t-t_1$ 
are not too small.

\subsection{Matrix elements: Hadronic mixing}

 The problem of determining hadronic matrix elements involved in mixing
-  for example in mixing of glueball and scalar meson, in string
breaking or in flavour-singlet meson studies -  is much less
straightforward. In full QCD, mass eigenstates can be determined
directly and one is  able, much as in experiment, to determine the
masses of the resulting  mixed states. In quenched, or partially
quenched, studies it is possible to study mixing more directly by
evaluating  correlators between the different states involved. This is
the area we explore here.

 In lattice QCD with Euclidean time, the main factor is that the
lightest state  allowed will dominate the correlation at large time
separation $t$.  Thus in a study of glueball decay to two pions, for
example,  the two pion state will be lightest and dominate. In general 
this makes hadronic decays difficult to study~\cite{cmdecay}. As we
shall see, the  way forward is to restrict study to transitions in which
the energies are similar: on-shell transitions.  In the case of glueball
decay to two pseudoscalar mesons, this would imply  varying the quark
mass until the two pseudoscalar mesons had comparable energy to  the
glueball, as was arranged in a pioneering study of this~\cite{sexton}.

 Here we have in mind several situations where energies are comparable: 
(i) scalar glueball mixing with scalar mesons, (ii) $B$ $\bar{B}$ mixing
 with a static $Q \bar{Q}$ at separation $R$, (iii) mixing  of $s
\bar{s}$ pseudoscalar mesons with singlet $q \bar{q}$ pseudoscalar
mesons. In each case one system is fully treated (i.e. the gluonic
interactions in the scalar glueball and the static $Q \bar{Q}$ or the $q
\bar{q}$ state with sea quarks $q$ in the vacuum) while the  other has
valence quarks which are treated in a quenched (or partially quenched) 
approach. 

\begin{figure}[htb]

\epsfxsize=9.5cm\epsfbox{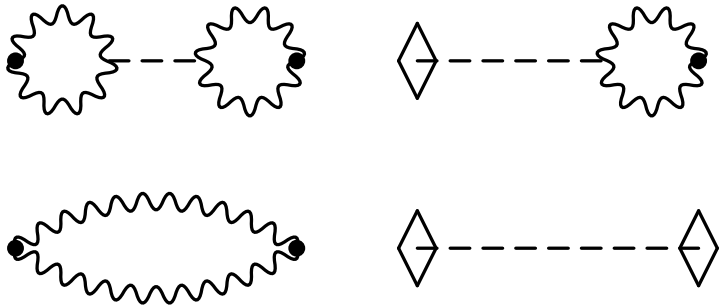}

 \caption{We illustrate correlations among scalar glueballs (created by
a closed Wilson loop) and scalar mesons made from quark-antiquark. 
Clockwise from the top left: the disconnected fermionic correlation
($D^{ss}$), the cross correlation of  a fermion loop with a Wilson loop
($C^{ms}$ or $H$), the correlation of Wilson loops ($C^{mm}$)  and the
connected fermionic correlation (${\cal C}^{ss}$).
 }
 \label{corr_fig}

\end{figure}

Our notation is that we consider $M$ states with masses $m_{\alpha}$ to
be fully described (that is to say they are either fermionic with those
quarks present in the vacuum or they are fully described in a quenched
theory as are glueballs or  potentials).  We also consider $M'$ states
with masses $s_{\beta}$ which are  quenched or partially quenched (i.e.
fermionic with  valence quarks which do not participate in the sea).
Here we assume that the spectrum of states is discrete in each case. To
be more specific, in one  application  one can think of the $m$ states
as glueballs and the  $s$ states as scalar mesons.  Some relevant correlations 
are illustrated in fig.~\ref{corr_fig}.

 Thus the $s$ states do not contribute to correlators unless an explicit
operator  creates or destroys them whereas the $m$ states can occur as
intermediate states  in any correlator with the correct quantum numbers.
 We shall treat the $s$ states as those given by the relevant connected
fermionic correlator. The disconnected contribution to any  correlator
will be included explicitly. The model used for these disconnected 
contributions may depend on the application. Assuming that the
disconnected  fermionic loops in the $s$ state to $s$ state correlator
are joined by  gauge links, one would expect transitions between these
$s$ states and  $m$ states to be the relevant description. In some
cases, however, such as pseudoscalar  mesons, the gluonic links between
the disconnected loops are expected to be short ranged and so are
treated as local. In this case  an explicit mixing coefficient between
disconnected $s$ states should be introduced.

 We will assume that transitions between states of mass $m_{\alpha}$ 
and $s_{\beta}$ are local and have strength $x_{\alpha \beta}$. The
goal  is usually to determine the transition strength between ground
states: $x_{00}$. Because the transition is hadronic, there is no 
explicit matrix element insertion. One must deduce the strength of the
transition  from a study of the two point correlators alone.

 Consider now the two point correlators at separation $t$ between
operators  creating either of these types of state. For each type of
creation or annihilation operator, we consider several different
operators labelled $i,j=1,N$ and $k,l=1,N'$ respectively (here we have
in mind different smearing or fuzzing prescriptions). Then we have a
description in terms  of transfer matrix eigenstates as 
 \be
 C_{ij}^{mm} (t)=\sum_{\alpha=0}^{M-1} c_i^{\alpha} e^{-m_{\alpha}t}
c_j^{\alpha}
 \ee
 \be
 C_{ik}^{ms} (t)=\sum_{t_1=0}^t \sum_{\alpha=0}^{M-1} \sum_{\beta=0}^{M'-1}
c_i^{\alpha} e^{-m_{\alpha}t_1} x_{\alpha \beta} e^{-s_{\beta}(t-t_1)}
d_k^{\beta}
 \ee
 There will also be additional terms with $t \to T-t$ for a lattice
periodic  in time with extent $T$ - these we do not write explicitly. 
Now for the quenched $s$ correlation one can separate it into connected 
and disconnected fermion contributions

 \be
 C_{kl}^{ss} (t)={\cal C}_{kl}^{ss}(t) + D_{kl}^{ss}(t)
 \ee
 with
 \be 
 {\cal C}_{kl}^{ss} (t)=\sum_{\beta=0}^{M'-1} d_k^{\beta}
e^{-s_{\beta}t} d_l^{\beta}
  \ee
 and 
 \begin{equation} D_{kl}^{ss}(t) =
   \sum_{t_1=0}^t   \sum_{\beta=0}^{M'-1} \sum_{\beta'=0}^{M'-1} 
   \sum_{t_2=t_1}^t \sum_{\alpha=0}^{M-1}
 d_k^{\beta} 
e^{-s_{\beta}t_1} x_{\alpha \beta} e^{-m_{\alpha}(t_2-t_1)} x_{\alpha
\beta'} e^{-s_{\beta'}(t-t_2)} d_l^{\beta'} 
 \end{equation}
 One might also include contributions to $D^{ss}$ coming from a direct 
(i.e. not via an $m$ state) transition at $t_1$  from state $s_{\beta}$
to $s_{\beta'}$ with mixing strength $y_{\beta \beta'}$. As discussed
above, this is appropriate  for a discussion of pseudoscalar meson
mixing, for example. In general, if  propagation of the states of mass
$m_{\alpha}$  over zero time steps is included, then the above  formula
does include contributions giving the effect of such direct transitions.

 The main problem with extracting the mixing matrix elements $x_{00}$ is
in removing the excited state contributions. Unlike in the case of the 
explicit insertion at $t_1$, here we have no dependence on $t_1$ (and
$t_2$) in the observables. This implies that $t_1$ values  from 0 to $t$
will be allowed so that there will be excited state contributions to the 
correlator which are not suppressed (since $e^{-(m_1-m_0) t_1}$ with
$t_1=0$ is large, for example).

 The cleanest way to circumvent this is in the case when the masses are
equal,  i.e. $m_0=s_0$. In this case,  called the on-shell case, the
mixing observable  $C_{ik}^{ms}(t)$  will have  a contribution from the
sum over $t_1$ of $c_i^0 x_{00} t e^{-m_0 t} d_k^0 $  from the ground
states whereas the dominant excited state contributions behave as 
$c_i^0 x_{01} e^{-m_0 t} d_k^1$ and will be suppressed  by a relative
factor of $1/t$. This is a much smaller relative suppression than  the
factor of $e^{-(m_1-m_0)t}$ which applies to two point correlators, but 
it is  sufficient to remove excited state contributions.  This  implies
that $x_{00}$ can in principle be extracted  unambiguously. Likewise
$D_{kl}^{ss}(t)$ has a contribution of $d_k^0 x_{00} t^2 e^{-m_0 t}
x_{00} d_l^0 /2$ from the ground states which also dominates, by a
factor of $t$,  any excited state contributions. This gives a further
cross-check since  $x_{00}$ can be determined in these two independent
ways. 

 The analysis when $m_0 \ne s_0$ is more tricky. The only
$t$-dependences that  can be observed will be of the form $e^{-m_\alpha
t}$ and $e^{-s_\beta t}$  (and also $t e^{-s_\beta t}$ for $D^{ss}(t)
)$. Then assuming that  the $M,\ M'$ state fits to $C_{ij}^{mm}(t)$ and
${\cal C}_{kl}^{ss}(t)$ yield the  masses $m_\alpha$, $s_\beta$ and
couplings $c_i^\alpha$ and $d_k^\beta$ of both ground  states and
excited states, the $t$ dependence of $C_{ik}^{ms}$ and $D_{kl}^{ss}$ 
are available to determine $x_{\alpha \beta}$. In principle there are 
enough such independent $t$-dependences to determine the mixing
parameters, given sufficiently precise data. 

 One way to see this is to use the variational formalism with
eigenvectors $u^\alpha_i$ obtained from $C_{ij}^{mm}(t)$ and 
$w^\beta_i$ obtained from ${\cal C}_{kl}^{ss}(t)$ using  time values 
of $t$ and $t-1$. Then we can project into this variational basis
 \be
  H_{\alpha \beta}(t)= u^\alpha_i C_{ik}^{ms}(t) w^\beta_k
 \ee
 If the variational basis corresponds to the exact spectrum then we
would have,  using continuum evaluation of the sum over $t_1$ as would
be appropriate  for a very small lattice spacing, 
 \be
    H_{\alpha \beta}(t)= x_{\alpha \beta} 
   {e^{-m_\alpha t} - e^{-s_\beta t} \over s_\beta - m_\alpha }
 \ee
 from which $x_{\alpha \beta}$ can be extracted. In general, however, 
the variational basis does not correspond to the exact spectrum. A fit
to  the $t$-dependence then is needed. Provided enough operators $N$ and
$N'$  are used (namely $N > M,\ N' > M'$), there is sufficient
information to extract the parameters $x_{\alpha \beta}$ in principle. A
similar variational analysis  of $D^{ss}_{kl}(t)$ is also possible and
this gives another way to determine constraints on  the parameters
$x_{\alpha \beta}$.
 If $\beta = \beta'$ then, using continuum evaluation of the sums over $t_1$ 
and $t_2$,

 \be
    D_{\beta \beta'}(t)= 
\sum_\alpha     x_{\alpha \beta}   x_{\alpha \beta'} 
   { e^{-m_\alpha t} - e^{-s_\beta t}( 1+(s_\beta -m_\alpha)t)
   \over (s_\beta - m_\alpha)^2 } 
 \ee

 while if $\beta \ne \beta'$, then
 \begin{eqnarray} 
&&\mbox{}    D_{\beta \beta'}(t)=  
 \sum_\alpha     x_{\alpha \beta}   x_{\alpha \beta'} 
 \left(  { e^{-m_\alpha t}    \over (s_\beta - m_\alpha)  
                                        (s_{\beta'} - m_\alpha) } \right.
 \nonumber \\
&&\mbox{} \left.
+   { e^{-s_\beta t}    \over (s_\beta - m_\alpha)  (s_\beta - s_{\beta'}) }
+   { e^{-s_{\beta'} t} \over (s_{\beta'}-m_\alpha) (s_{\beta'} - s_\beta) }
 \right) 
 \end{eqnarray}

 Note that this approach assumes that an accurate description of the 
diagonal two-point correlators exists which is valid down to $t=0$. 
This is in practice hard to achieve. In particular the $t=0$ correlator
is  often quite different (sometimes even in sign) from the $t > 0$
correlators.  We now address the possibility of excited states that
contribute only to $t=0$ since these are needed to cope with this 
data. 


 Let us explore this situation with  $m_0 \ne s_0$ in a simple example. We
will assume precise data are available  for the correlations at all $t$
values and that, by an optimal choice  of operators, the diagonal
correlations  $C^{mm}$ and ${\cal C}^{ss}$ are described exactly by one state
for $t \ge 1$ and so have an additional excited state contribution of
the form 
 \be
   C^{mm}(t)=c^0 e^{-m_0 t}c^0 +c^1 \delta_{t0} c^1
  \ee \be
   {\cal C}^{ss}(t)=d^0 e^{-s_0 t}d^0 +d^1  \delta_{t0} d^1
 \ee
 Here we suppress the operator labels ($i,j,k$ etc) since we are
considering the case that  an optimum combination of them has already
been taken to isolate the ground state (i.e. just taking $\alpha=0$ and
$\beta=0$ above). Then the cross correlation has the form for $t > 0$,
including the excited state  contributions from $t_1=0$ and $t_1=t$,
 \be
  H(t)= \sum_{t_1} c^0 e^{-m_0 t_1}x_{00} e^{-s_0 (t-t_1)} d^0 +c^1 x_{10}
e^{-s_0 t} d^0 + c^0 e^{-m_0 t} x_{01} d^1
 \ee
   Now by completing the
sum over $t_1$ as a discrete sum,  $H(t)$ can be expressed as  
 \be
    H= c^0d^0 e^{-m_0 t} (A+ B \lambda^t) 
 \ee
 with $\lambda=e^{-(s_0-m_0) }$,
 \be   A={x_{00} \over 1-\lambda}  +{d^1 x_{01} \over d^0}
 \ee  and 
 \be B=-{\lambda x_{00} \over 1-\lambda}  +{c^1 x_{10} \over c^0}
 \ee
 Here $d^0,\ d^1,\ c^0,\ c^1$ and $\lambda$ are known in principle but
the  mixing parameters $x_{00}$, $x_{01}$ and $x_{10}$ are to be
determined.
 With perfectly precise data for $H$, only the coefficients $A$ and $B$
can be determined, hence the three mixing parameters cannot be
independently determined.  Thus $x_{00}$ cannot be determined, even
in principle.  The exception to this is when  $\lambda \to 1$, since the
contribution of $x_{00}$ is then greatly enhanced. In  fact in this case
$x_{00}$ can be read off from the coefficient of the linear term in $t$
in $H$ as discussed above.

 Now when data are available for the  correlation $D$ then additional 
constraints exist. For our example this is 
 \be
  D(t)=d^0 d^0  e^{-m_0 t} (X+Y\lambda^t+Zt\lambda^t )
 \ee
 with 
 \be 
 X=({x_{00} \over 1-\lambda})^2+ (d^1 x_{01} /d^0)^2+{2x_{00} x_{01} d^1
\over (1-\lambda)d^0}
 \ee
 \be 
 Y=-\lambda(2-\lambda)({x_{00} \over 1-\lambda})^2 -{2x_{00} x_{01} d_1
\lambda \over (1-\lambda)d^0} +2x_{10}x_{11}d_1/d^0+x_{10}^2
 \ee  and 
 \be
Z=-\lambda x_{00}^2/(1-\lambda)+x_{10}^2
 \ee


Here $x_{11}$ is a new mixing:  between the excited states in both sectors.
We now have three additional constraints ($X$, $Y$ and $Z$)
given accurate data  for $D(t)$, with only one additional parameter.
Thus in the case of this simple  model, the measurable quantities
overdetermine the mixing parameters. Again as $\lambda \to 1$, this
expression  simplifies and the coefficient of $t^2$ in $D(t)$ gives
$x_{00}^2$ directly.

 One way to check that ground state contributions dominate is to extract
 $x_{00}$ from $H$ and $D$ at several $t$ values neglecting excited
states and check for consistency. So the relevant expressions will be
 \be
  x_H(t)=  {H(t) \over \sqrt{C^{mm}(t) {\cal C}^{ss}(t)} }
{ \lambda^{t/2} \over 1 + \lambda + \dots + \lambda^t}
 \label{heq}
 \ee  and
 \be
  x_D(t)= \sqrt{D(t) \over  {\cal C}^{ss}(t) }  { \lambda^{t/2} \over
 \sqrt{1 + 2\lambda + \dots + (t+1)\lambda^t} }
 \label{deq}
 \ee

 In our application below $\lambda = 0.64$ for strange quarks, so the 
enhancement of the ground state mixing ($x_{00}$) by factors of
$1/(1-\lambda)$ is not very big. However, we use information from both
$H$ and $D$ which does provide a cross check in principle of our
assumption that the excited state contributions (such as $x_{01}$ and
$x_{10}$  in the above example) are negligible.

 In summary, when we have an on-shell hadronic transition we can extract
the  mixing matrix element with a power suppression in $t$ of excited
states compared to  the exponential suppression which applies to
extracting masses. When the masses  are not equal, excited states can
still be removed in principle if a precise study  is made of both
correlations of type $H$ and $D$. Note that in practice, if the 
difference in masses is significant, the extraction of the mixing
strength will be  very difficult.

\section{Lattices}

 We choose to study sea quark effects using the configurations  with
$N_f=2$ at $\beta=5.2$ with $C_{SW}=1.76$ from UKQCD~\cite{ukqcd}.
 Two spatial lattice sizes are available ($12^3$ and $16^3$) so that 
finite size effects can be explored. We use the three lightest  sea
quark masses available and  we use valence quark masses equal to the 
sea-quark mass. The lattice information is summarised in  Table~1. Local
and non-local meson operators were used  with fuzzing radius 2 with 5
iterative levels with coefficient 2.5.

We also use for comparison a quenched lattice at $\beta=5.7$ of size 
$12^3 24$ with valence fermions of $\kappa=0.14077$ (approximately
strange mass)and $\kappa=0.13843$ (approximately twice strange mass)
with $C_{SW}=1.57$,  as studied previously~\cite{shan}.

 In order to improve the statistics we measure the disconnected diagrams
on  configurations separated by  less trajectories than for the
connected correlators as shown in Table~2. Even though there may be some
autocorrelation among these measurements  separated by less
trajectories, we find that this approach does allow the statistical
error to be reduced. Indeed this is the approach that was used in
glueball studies, where  the measurement time is very small so one might
as well measure almost every  configuration - indeed we follow this
approach here when considering glueballs, as discussed later.

\begin{table}[t]
\footnotesize
\begin{tabular}{llllll}
\hline
 $L_s$& $\kappa$ & $r_0/a$ & $m_{\pi}a$ & $m_{\pi}/m_{\rho}$ \\

\hline  
12&0.1390 & 3.05 & 0.707(5) & 0.78 \\
16&0.1390 & 3.03 & 0.701(6) & 0.78   \\

12&0.1395 & 3.44 & 0.558(8) & 0.71   \\
16&0.1395 & 3.44 & 0.564(4) & 0.72   \\

12&0.1398 & 3.65 & 0.476(16) & 0.67  \\
16&0.1398 & 3.65 & 0.468(5) & 0.67  \\

12&0.13843& 2.94 & 0.736(2) & 0.78 \\
12&0.14077& 2.94 & 0.529(2) & 0.65 \\

\hline

\end{tabular}
 \caption{Properties of the $N_f=2$ lattices from UKQCD~\cite{ukqcd}
with $\beta=5.2$ and $C_{SW}=1.76$ and (last two rows) $N_f=0$
lattices~\cite{shan} with $\beta=5.7$ and $C_{SW}=1.57$. }

 \label{latts.tab}

\end{table}

\begin{table}[t]
\footnotesize
\begin{tabular}{llllll}
\hline
 $L_s$& $\kappa$ & $N_S$&connected& disconnected  & glueball\\
     &           &      &configs &   configs& configs\\

\hline  
12&0.1390 &24& 151&301 & 301 \\
16&0.1390 &200& 90&94 & 390 \\

12&0.1395 &24& 121&253 & 505 \\
16&0.1395 &48& 100&106 & 424 \\

12&0.1398 &24& 98&169 & 170 \\
16&0.1398 &48&69& 75 & 298 \\

12&0.13843&24&482&100&100 \\
12&0.14077&24&482&100&100 \\

\hline

\end{tabular}
 \caption{Statistics of connected, disconnected and glueball
calculations. For the fermionic disconnected correlation, the variance
reduction  methods with $N_S$ samples were different for $L_s=12$ and 16
as described in the appendix.}

 \label{stats.tab}

\end{table}

\section{Scalar mesons}

  Within the quenched approximation,  there will be two distinct types
of scalar meson:  $q \bar{q}$ mesons and scalar glueballs. In full QCD,
these  two types of state will mix, resulting in the observed experimental
spectrum of  scalar mesons. As an aid to disentangling this experimental
situation, we here  explore the lattice predictions for scalar mesons.
As a first step we evaluate the mixing  matrix elements in the quenched
approximation. This has been  explored~\cite{lw} previously and here we
discuss the problems associated  with determining such hadronic mixing
on the lattice.

\subsection{Quenched lattice results}

\begin{figure}[htb]

\epsfxsize=9.5cm\epsfbox{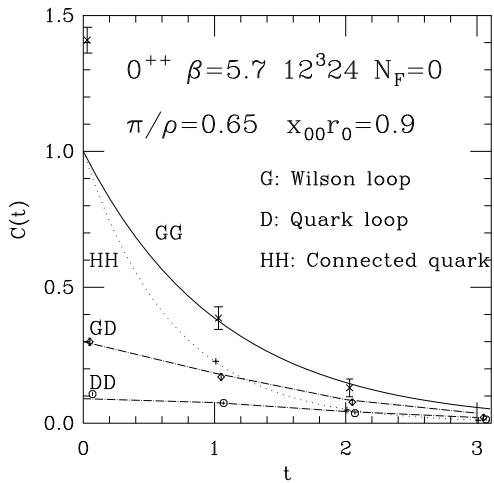}

 \caption{Quenched scalar correlations with quark masses approximately
strange. Here HH is the connected mesonic correlation  (${\cal
C}^{ss}$), GG is the glueball correlation  ($C^{mm}$), DD is the
disconnected mesonic correlation  ($D$) and GD is the cross correlation
between glueball and meson operators ($H$).  Lattice results are
illustrated  for one glueball operator and one (local) meson operator.
The curves are  a simple mixing model, as described in the text.
 }
 \label{mixing}

\end{figure}

\begin{figure}[htb]

\epsfxsize=9.5cm\epsfbox{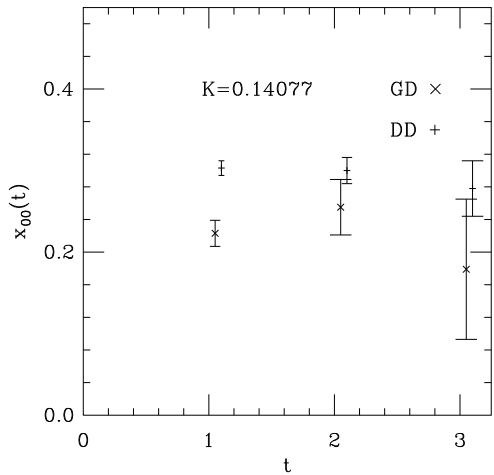}

 \caption{The mixing coefficient $x_{00}$ is determined in lattice units
 with $a^{-1} \approx 1$ GeV, at each $t$ value,  from  quenched scalar
correlations with quark masses approximately strange:  DD is the
disconnected mesonic correlation  ($D$) and GD is the cross correlation
between glueball and meson operators ($H$). 
 }
 \label{mixing_ratio}

\end{figure}

\begin{figure}[htb]

\epsfxsize=9.5cm\epsfbox{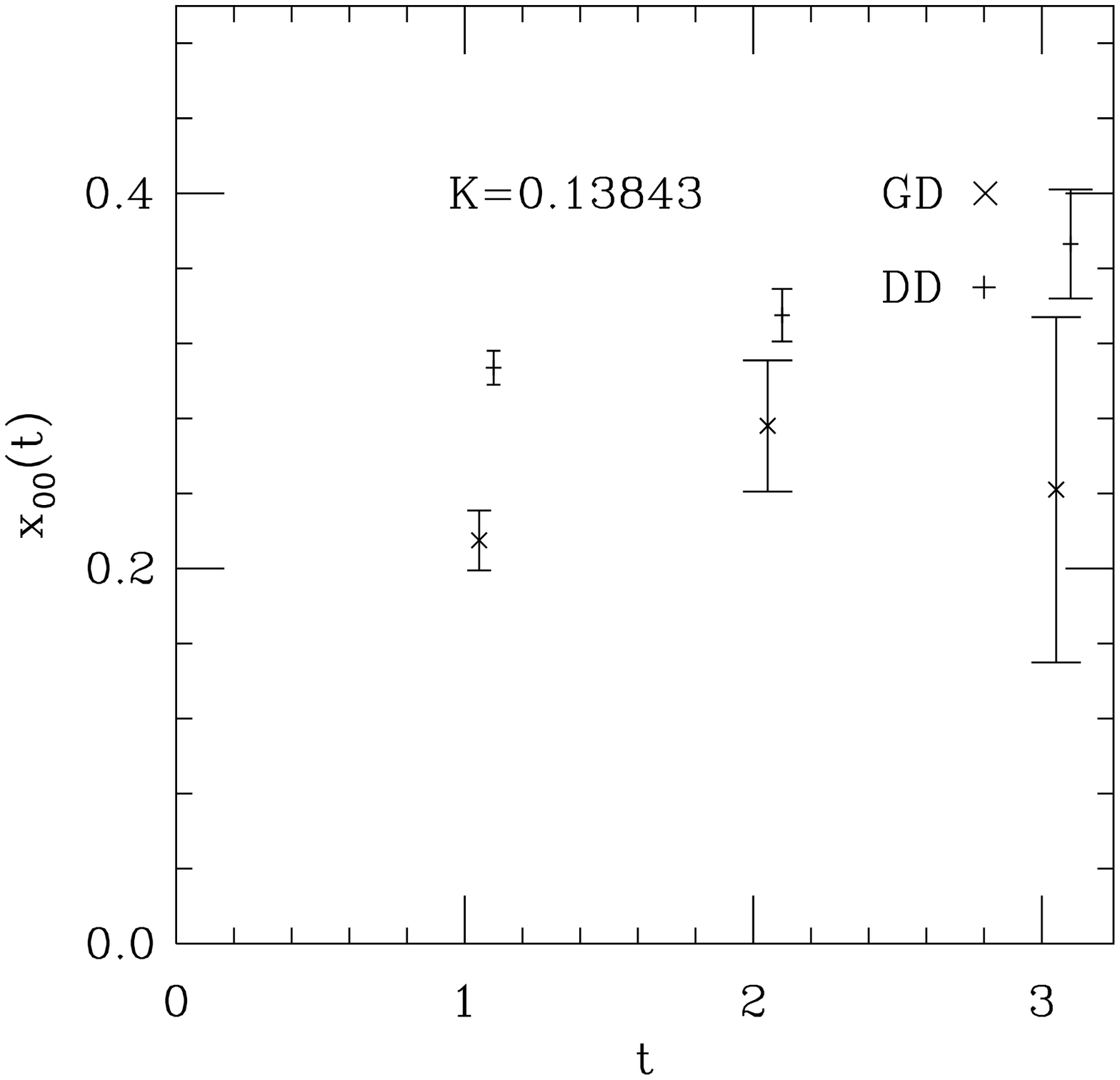}

 \caption{As fig.~\ref{mixing_ratio} but with  quenched scalar correlations
with quark masses approximately twice strange.
 }
 \label{mixing_ratio2}

\end{figure}

 We explored this mixing in the quenched approximation (see Tables~1 and
2) using  SW-clover valence quarks of two different masses~\cite{shan}. 
The zero-momentum glueball operators  were measured at every time-slice
in the usual way~\cite{mt} and the disconnected  quark loops are
measured as described in the appendix, namely with sufficient stochastic
samples that  no significant error arises from the stochastic algorithm.
 The connected quark correlators were taken from previous
measurements~\cite{shan}. Since the scalar meson or glueball has vacuum
quantum numbers,  we subtract the vacuum contribution in the other types
of  correlation we measure.  Our results for all of these types of
correlation are illustrated in fig.~\ref{mixing} for the case of one
choice  of glueball operator and one (local) mesonic operator at  our
lighter  quark mass.

 From the connected quark correlator ${\cal C}^{ss}$, at hopping
parameters $\kappa=0.14077$ and  $0.13843$, we find a  scalar $q\bar{q}$
masses of 1.39(5), 1.36(2) respectively in lattice units, fitting local
and  two types of smeared operator to  one state in the $t$ range 2 to
8. Note that the mass ordering is  not as expected (namely meson with
lighter quarks being lighter) but the errors are large enough to cover
near equality.  This meson mass value  is somewhat larger than that
reported~\cite{lw} at the same $\beta$ value but using Wilson quarks of
mass corresponding to strange (ie our $\kappa=0.14077$), namely 1.29(2).
This discrepancy is not surprising since the SW-clover formalism we use
has improved control of order $a$ effects compared to the Wilson
discretisation. 

 For the glueball mass, which is of course independent of fermion
formalism in quenched studies,  we use the higher statistics
result~\cite{lw} of 0.95(2)  in lattice units obtained for  $t\ge 2$.
Our result for the glueball correlator ($C^{mm}(t)$) is consistent with
a single exponential with this mass for  $t \ge 1$.  Since our glueball
correlator has  large errors for $t > 1$,  in evaluating the expressions
shown in  figs.~\ref{mixing_ratio} and \ref{mixing_ratio2} we use our
measured glueball correlations  at $t=1$ but for $t>1$ we assume the
glueball correlation has  the mass dependence given by $ma=0.95$ as
found in higher statistics studies. As shown in fig.~\ref{gbr0}, this
glueball mass  lies below the continuum extrapolation because of order
$a^2$ lattice artefacts.
 To convert to physical units, we use $r_0/a=2.94$ and then 
conventionally $r_0 \approx 0.5$ fm, so $a^{-1} \approx 1.1$ GeV.

 Then given these mass values, one can attempt to describe the
disconnected  and cross correlations ($D$ and $H$) in terms of the one
free  parameter, the mixing strength $x_{00}$.  The errors on our
determinations of  these  correlations are quite large: 25\% for $D$ and
50\% for $H$ at $t=3$. We use local fermionic operators in these
comparisons since the smearings used in the determination of the
connected and disconnected fermionic correlator were different for
historical reasons. 

The curves shown in  fig.~\ref{mixing} are from the lattice model
described above with ground state contribution  only and  with
$x_{00}a=0.3$. This is seen to  give a reasonable overall description
for $1 \ge t \ge 3$. Alternatively,  in fig.~\ref{mixing_ratio},
\ref{mixing_ratio2}  we give the value of the mixing parameter $x_{00}$ 
for each $t$-value obtained from taking the data on $H$ and $D$
respectively  and assuming the lattice mixing  expressions of the
previous section (eqs \ref{heq}, \ref{deq}) with no excited state
contributions. 
 We find for each quark mass that $x_{00}a \approx 0.3$. 

  Since our results for the diagonal correlations of glueball and meson
operators  ($C^{mm}$ and ${\cal C}^{ss}$) are only reasonably 
approximated by the ground state for  $t_g \ge 1$ and $t_q \ge1$, as
discussed above,  the mixing should be studied from  correlations with
$t \gg 2$. As shown in fig.~\ref{mixing_ratio},  we find that the signal
 becomes very noisy by $t=3$ already. With 100 times the statistics we
would be able to  determine the $t=4$ mixing correlations ($D$ and $H$)
to 10\%. This implies that much larger  data sets (number of gauge
configurations about 10000)  are needed to give a more definitive 
answer to the mixing in this case. Even then since there is only a power
suppression of excited states, one would need precise data to large $t$
to have small systematic errors  from excited state contributions. 

 The consistency of the determinations of $x_{00}$ from different $t$ values
and  different quantities does, however,   act as a cross check that our
results are consistent with the assumption that a single ground state
dominates.
 Because of the lack of control of excited  state contributions, we can
only quote the statistical error on  the mixing $x_{00}$. Assuming no
excited state contamination and taking $t \ge 2$, we obtain
$x_{00}a=0.26(4)$ for  strange quarks, corresponding to  $x_{00} r_0=
0.76(12)$. This is the quenched result for one flavour and we see no
sign of any significant difference in $x_{00}$ as we  vary the quark
mass since we have $x_{00}a= 0.32(4)$ for heavier quarks.

A previous work~\cite{lw} has studied glueball mixing with a scalar
meson in the quenched approximation. They used several $\beta$ values,
Wilson fermions and concentrated on the cross correlation $H$ to 
determine $x_{00}$. At $\beta=5.7$ and for quark mass near strange, we
can make a direct comparison, bearing in mind  that the order $a$
corrections are significant at such a coarse lattice spacing  and will
be different for Wilson and SW-clover fermion formalisms, indeed  the
SW-clover formalism we use is focussed on removing these order $a$
effects.

 At $\beta=5.7$ they have $t_q=2$ and $t_g=2$ and they determine their 
mixing coefficient from data for $H$ with $t \ge 2$ by assuming  no
excited state contributions to $H$.  They do not consider data on $D$.
Note that, as discussed above, from measurement of  $H$ alone, it is
impossible in principle to confirm that excited state contributions are 
absent. Their quoted result for strange quarks  is $x_{00}a=0.211(16)$.
This is broadly compatible with our estimate of $x_{00}a=0.26(4)$  
bearing in mind that different  fermion discretisations were used.

 At larger $\beta$ (up to 6.2)  their results for the mixing are that, 
at the strange quark mass, the mixing tends to a very small value in the
continuum limit. The   situation concerning excited state contamination
is even worse at larger $\beta$ since they find $t_q=6$ and $t_g=4$ at
$\beta=6.2$, whereas they still fit $H$ for  $t \ge 2$. Thus their mixing
estimates at larger $\beta$ are even more  susceptible to excited state
contamination. Moreover they do not make use of the disconnected
correlator $D$ to constrain their assumptions further.

We find a significant mixing at coarse lattice spacings using a fermion
formalism that  has been shown to have reduced order $a$ corrections.
This implies that we would expect  a substantial mixing in the continuum
if order $a^2$ corrections were also to be small.  This is in contrast
to  the conclusion~\cite{lw} that the $a$ dependence of the  mixing (in
physical units) is such that the continuum mixing is very small. We
conclude that we find evidence for a mixing strength $x_{00}
r_0=0.76(12)$ with one flavour of quarks of strange mass at our lattice
spacing.

\subsection{Full QCD scalar mesons}

 To compare with our results using dynamical sea quarks with $N_f=2$, we
estimate  the effects of our quenched determination of the mixing if
applied to that case. For this estimate we take $x_{00}a=0.3$ for one 
quark flavour. Then for quarks of strange mass  our unmixed states with 
 $m_0 a= 0.95$ and $s_0 a= 1.39$ will be mixed by off diagonal element 
$x_{00}a=0.3 \sqrt{N_f}=0.42$ giving a mass mixing matrix:

 \begin{math} 
 \left( \begin{array}[h]{cc}
    am_0  &   \sqrt{2} ax_{00}  \\
    \sqrt{2} ax_{00}  &   as_0 \\
\end{array} \right) =
 \left( \begin{array}[h]{cc}
    0.95  &    0.42 \\
    0.42  &   1.39 \\
\end{array} \right)
 \end{math}

\noindent  which gives mass eigenstates pushed apart  to $ma=0.69$ and
1.65 (and for the heavier quarks with the same mixing strength then 0.95
and 1.36  will be mixed  to 0.68 and 1.63). 
 Thus for strange quarks, the lightest scalar meson would be reduced in
mass  by approximately 0.24 in lattice units.  If our  quenched mixing
strength were be to applied to the scalar mass matrix, it results in a
downward shift for $N_f=2$ of the lattice glueball  mass by 25\%.

 We now consider the sea-quark case explicitly, where the mixing will be
 observed directly from the resulting mass values. What we can measure
in that case  is the non-singlet mass and the ground and excited states
in the flavour singlet sector. Based on the results from the quenched
study, we would expect in the flavour singlet sector that the ground
state mass  lies  considerable below the flavour non-singlet mass and
the first excited state mass is slightly above the flavour non-singlet
mass.

We use 4 scalar meson operators (i) closed Wilson loops (glueball
operators) of  two different sizes (Teper-smeared~\cite{mt}) and (ii) $q
\bar{q}$ operators which are local and separated by fuzzed links. For
the fermionic correlations, we include the connected and disconnected
contributions as given by the Wick formalism. We evaluate the  $4 \times
4$ matrix of correlations. Since different  numbers of gauge
configurations are analysed for different operators, where necessary, we
 average results from nearby gauge configurations to have a consistent
bootstrap sample for error analysis.

We should also consider two particle states (two pseudoscalar mesons 
with angular momentum 1, for example) which can mix with scalar mesons
and glueballs. On a lattice the lightest such  state with overall
momentum zero will have energy $2a(m_{\pi}^2 +4 {\pi}^2/L_s^2)^{1/2}$
which is at 1.53 for the case of $L_s=12$ and $\kappa=0.1395$ - this
will apply  to the flavour singlet case.  For the flavour non-singlet 
scalar meson then the $\pi \eta$ mode will be the lightest and this will be
even heavier. These two particle energies are sufficiently heavy that we
shall ignore these states in our present work. They will, however,
become important as the quark mass is reduced  and the lattice size is
increased.

The glueball  and fermionic singlet correlations have a vacuum
contribution.   For the glueball fits  we deal with this by using 2
state fits with one state constrained  to have mass=0 (namely the
vacuum). In such fits it is important to use correlated fits to have a 
meaningful expression for $\chi^2$. Since we are fitting to many (up to
60)  different types of  data (different operators at source/sink and
different $t$ values), we use the technique of retaining  exactly the
$N_e$  largest  eigenvalues of the correlation among the data set and
setting the remaining eigenvalues  equal~\cite{mm}.  This avoids
spurious correlations being induced because of our limited sample size. 
We use $N_e=10,8$ for $L_s=12,16$ respectively.
 For our fits to the full $4 \times 4$ matrix of  singlet correlations,
since the number of gauge samples is different for different 
observables which makes the vacuum contribution depend on the
observable, it is preferable to subtract the vacuum contribution  and
fit the resulting connected correlation.  The errors quoted on the fits
are  statistical from bootstrap analysis and do not include systematic
errors from varying the  fit range or fit function: these are at least
comparable in size.

\begin{figure}[htb]

\epsfxsize=9.5cm\epsfbox{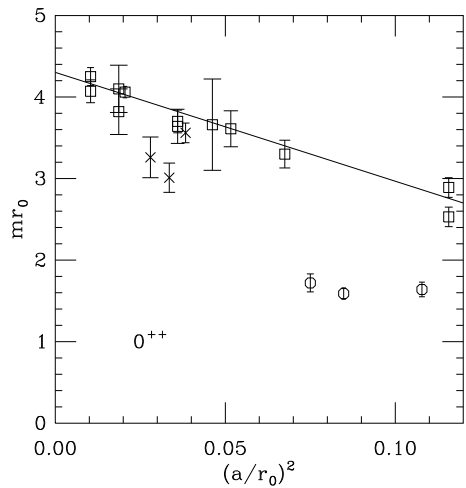}

 \caption{ The scalar mass versus $a^2$. The quenched
results~\cite{DForc,mt,ukqcdg,gf11} are for the scalar  glueball and are
shown by  boxes. The results from $N_f=2$ flavours of sea quark are from
glueballs~\cite{SESAM} (crosses  from SESAM) and  the lightest flavour
singlet scalar we find here (circles).
 }
 \label{gbr0}

\end{figure}

 We can now measure directly the scalar spectrum for $N_f=2$ to explore
this. Results are shown in Table~\ref{mfits}.  We obtain the favour non-singlet
mass ($m_{NS}$) from a two state fit from $t=2$ to 7 to the  $2 \times
2$ matrix of connected fermionic correlators. For the singlet, we now
use both glueball and $q \bar{q}$ operators (vacuum subracted)  and find
an acceptable fit with 1 state to  the $4 \times 4$ matrix of
correlations for $t$ from 2 to 7: the results are shown as $m_{FS}a$  in
Table~\ref{mfits}. We find that the mass obtained from fitting only the $2 \times
2$ matrix of  glueball correlations ($m_{GB}$) is consistent with the 
full fit, as it should be. Moreover,  we see a surprisingly low  scalar
mass - as emphasised in fig.~\ref{gbr0} which compares with quenched
results and the SESAM $N_f=2$ values~\cite{SESAM}. 

It would be interesting, as discussed above, to  obtain the excited
state mass in the flavour singlet sector. We expect this above $m_{NS}$
at  around $am'=1.4$. We have used the variational  method for $t=1,2$
to extract the two lightest mass eigenstates from our $4 \times 4$
matrix  of vacuum-subracted correlators. This variational  ground state
mass  (0.44(1) for $\kappa=0.1395$) agrees quite well with the fitted
value shown in  Table~\ref{mfits}, as expected. The next state is poorly
determined  although for the case with best statistics ($L_s=12$ and
$\kappa=0.1395$) we  find $am'=1.27$ which is close to $am_{NS}$. It
will be interesting to explore this further with higher statistics. 
Note that it is difficult  to determine this mass   since the signal is
swamped by that of two lighter states (the vacuum at $m=0$ and the 
ground state at $am \approx 0.5$).

\begin{table}[tb]
\footnotesize
\begin{tabular}{llllll}
\hline
$L_s$& $\kappa$ & $m_{GB}a$ & $m_{FS}a$ & $m_{NS}a$ \\
\hline  
12&0.1390 & 0.40(6) &0.54(3)  &1.23(4) \\
16&0.1390 & 0.53(7) &0.47(3)  &1.19(5)\\
12&0.1395 & 0.49(4) &0.46(2)  &1.23(4)\\
16&0.1395 & 0.70(9) &0.75(4) &1.18(8)\\
12&0.1398 & 0.48(10)&0.47(3)  &1.00(5)\\
16&0.1398 & 0.58(8) &0.66(4)  &0.99(6)\\
\hline
\end{tabular}
 \caption{Ground state scalar masses from fits to the glueball sector
(GB); the whole  flavour non-singlet sector (FS: glueball and fermionic
operators) and to the  fermionic non-singlet sector (NS). The errors quoted 
are statistical only. }
 \label{mfits}

\end{table}

We do expect a relatively light flavour-singlet scalar mass because of
mixing  effects as described above which would reduce the 
mass by 25\%. This could explain in part our low
scalar mass but  other explanations are also worth exploring. For
example the order $a^2$  corrections might be anomalously large for our
lattice implementation (e.g. twice as large as in the quenched
Wilson case).

 Another possible explanation of the light flavour-singlet scalar mass
we find  would be a partial restoration of chiral symmetry  in a finite
volume, however, this would be expected only for very light quark masses. 
We do find that our flavour-singlet scalar meson masses are {\em
lighter} than the  pseudoscalar non-singlet (pion) mass for $L_s=12$ for
the range of  sea quark masses considered here.  The spatial size  with
$L_s=12$ is 1.7 fm and no evidence of finite size effects  was seen here
in a study of flavour non-singlet correlators~\cite{ukqcd}. Indeed we
see no significant  sign of any spatial size dependence in the
non-singlet scalar masses reported in Table~\ref{mfits}. 

 To explore this further,  we have made a study of flavour singlet
correlators on  $16^3$ spatial lattices to check directly for  finite
size effects and the results are  presented in Table~\ref{mfits}. The
signal to noise from zero momentum correlations is  worse for the larger
volume for glueball and disconnected correlations.  Also we find that
the  excited state contributions are relatively stronger for our
operators. Thus the systematic fit errors for $L_s=16$ are also
considerably larger than those for the smaller volume. Even though  the
signal from the larger spatial volume is relatively poorly determined, 
we do see some evidence (at the $2\sigma$ level if the sytematic errors
are taken as comparable to the statistical ones) of a  higher scalar
mass (for $m_{GB}$ and $m_{FS}$) on the larger spatial lattice. In order
to explore larger $L_s$ values, the signal to noise can be improved by 
considering  non-zero momentum correlators, as is the case for glueball
studies~\cite{mt}. 

One conclusion is that it would be valuable  to use a finer lattice
spacing or an improved gauge discretisation  so that any suppression of
the glueball  mass by order $a^2$ effects would be reduced. This would
increase the glueball mass and hence reduce the magnitude of the signal
we see, but  it would move the parameters into a region closer to
experiment.

\section{Conclusion}

 Hadronic mixing as exemplified by the glueball mixing  with
a scalar meson can be explored using lattice methods. In the quenched
approximation,  one can determine the mixing strengths although the
systematic errors in this  determination are large as we have discussed.
In studies with sea quarks, the mixed  spectrum itself is obtained which
gives complementary information.  

 In a preliminary study of this glueball mixing, we find a large mixing 
in a quenched study and, consistently, a large suppression of the mass
of the lightest scalar meson when sea quarks are included. These studies
are  computationally difficult and we have used a coarse lattice spacing,
albeit with an  improved (SW-clover) fermion formalism. It will be
necessary to extend  these  studies to smaller lattice spacing in order
to have more confidence in their relevance to the  phenomenological
situation.

\section{Appendix: Disconnected fermion loop evaluation}

 To evaluate disconnected quark loops with zero momentum, we need to sum
 over propagators from sources at each spatial location at a given time
slice. This  problem has been approached using stochastic source
methods~\cite{kuramashi,z2liu,sesam,roma-tv96}. Here we describe in more
detail the variance reduction techniques described
previously~\cite{pisa}. For a similar method see ref.~\cite{wilcox}.


 To study flavour singlet mesons, we need to consider quark loops  which
are disconnected (often called hairpins), namely evaluate $\rm{Tr}
\Gamma M^{-1}$ where the sum (trace) is over all space (for zero
momentum) at a given  time value and all colours and spins. Here $M$ is
the lattice fermion matrix, $\Gamma$ is a combination of the 
appropriate $\gamma$-matrix and a product of spatial gauge links if a
non-local (fuzzed) operator is used for the meson.

 Using a random volume source $\xi$ (where $\langle \xi^* \xi \rangle=1$
 for the same colour, spin and space-time component and zero otherwise)
then  solving $M \phi = \xi$, one can  evaluate unbiased estimates of
the propagator $M^{-1}_{xy}$  from $\langle \xi^*_y \phi_x \rangle$
where the average is over the $N_S$ samples  of the stochastic source.
The computational overhead of this method lies entirely in  the
inversion of $M$ to obtain $\phi$ from $\xi$ for each of the $N_S$
samples.

The drawback of this approach is that the variance on these estimates
can be very large, so that typically  hundreds of samples are needed. 
Here we present a method which succeeds in reducing this variance
substantially at rather small computational expense.

 The variance reduction  is based on expressing the fermion
matrix $M$ as
 \be
 M=C-D = C(1-C^{-1}D)=(1-DC^{-1})C
 \ee
 where $C$ is easy to invert, for example the SW-clover term which is
local  in space or in the Wilson case where one can chose $C=1$. Then we
have the exact identity 
 \begin{eqnarray}
  M^{-1} \!\!\!\!\! & = & \!\!\!\!C^{-1}  +C^{-1}D C^{-1} + \dots + 
               (C^{-1}D)^m C^{-1} \nonumber \\
 & +&  (C^{-1} D)^{n_1} M^{-1}(D C^{-1})^{n_2}  
 \label{mhit}
  \end{eqnarray}
      with $n_1+n_2=n=m+1$.

 Our strategy will be to evaluate the last term in this expression
stochastically and to evaluate the preceding terms as exactly as
possible. We will refer to these terms in our following discussion as
the stochastic and exact terms.
 If these exact terms can be evaluated precisely, then it is plausible
that the  stochastic term will contribute less variance to the overall
estimate of  $M^{-1}$ than in the $m=0$ case where there would be no
such exact terms. These exact terms can be evaluated either directly
(for example terms with odd powers of $D$ vanish in the evaluation  of a
local trace) or as a subsidiary  stochastic calculation with more
samples to achieve good precision and at relatively small computational 
overhead since no inversion is required.

 Since this approach is a variant of the hopping parameter  expansion,
it might be suspected that the convergence was poor since  at each
higher order  8 extra terms are present with coefficients which are of
order $\kappa \approx 1/8$. In our application, however, these 8 terms
contribute with random  strengths - like a random walk. So they have an
effective weight which is  more like $\sqrt{8}$ which is smaller. So we 
do find a reduced variance including more terms, but at the cost of some
extra  computation in evaluating the additional exact terms on the right
hand side  of eq.~\ref{mhit}.

 A special case of this ($n_1=n_2=2$) with Wilson fermions (for which
$C=1$  and the terms with up to 3 powers of $D$ vanish for $\rm{Tr}
M^{-1}$) employing Gaussian noise was used by the bermion
group~\cite{bermions} previously.

 Using the stochastic volume source, the variance reduced expression can
be rewritten (assuming $C$ is hermitian) as
 \be
 \sum_x \Gamma_{xy} M_{yx}^{-1} = \dots + 
     \langle (  (C^{-1}D^{\dag})^{n_2} \xi)_x^* 
 \Gamma_{xy} ( (C^{-1 } D)^{n_1} \phi )_y  \rangle
 \ee
 so the stochastic term may be evaluated as an average over  stochastic
samples $\xi$ after inversion to obtain $\phi=M^{-1} \xi$. In the
application of this  paper we take $\Gamma$ to have the Dirac structure
$I$ whereas in other applications~\cite{pisa,prl}  we consider $\gamma_5$
and $\gamma_5 \gamma_4$ also.

Only the even exact terms in the series when $\Gamma$ is local are
non-zero  and we calculate the $m=0$ and $2$ cases explicitly. For
hadron operators with fuzzed paths  of length $n_F$ the series starts at
$m=n_F$ (this we calculate explicitly) and then has alternate terms 
zero. The explicit calculations referred to  are rather cumbersome for
clover fermions, so in some cases we actually evaluate the simpler 
Wilson expression  and then evaluate the difference stochastically.  The
generic non-zero terms ($A$) in the series  were calculated
stochastically using  $4N_S$ Z2 noise samples $\xi$ using
 \be
  \sum_x \Gamma_{xy} A_{yx} = \Gamma_{xy} \langle \xi_x^* A_{yz} \xi_z \rangle
  \ee

 Taking $n_1 \approx n_2$ gives an averaging  over a smaller volume than
taking an asymmetric choice. We find that an asymmetric choice gives  a
smaller variance, presumably because it does involve averaging  over a
larger volume.   For different disconnected observables, the optimum
strategy is  not necessarily the same. In this work, we find a overall
good choice to be $n_1=0$, $n_2=16$ with Gaussian noise. This  results
in a reduction by a factor of 0.2 to 0.3 of the standard deviation of
the samples.


 Using larger values of $n_1$ and $n_2$ implies that the estimate of 
$M^{-1}$ is  very non-local, involving $\xi$ values up to $n_1$ and
$n_2$ away. To evaluate correlators between traces  at $t_1$ and $t_2$,
one must require that the samples of stochastic  volume source used in
the two cases are different so that there is no bias.  We use $N_S=24$
stochastic samples and this condition is readily implemented with
essentially no  loss of statistics.  This number of samples was chosen
to make the stochastic sampling error smaller  than  the intrinsic
variance from one time slice (for example at $L_s=12$ with
$\kappa=0.1395$, the  ratio of the stochastic sampling error to the
 standard deviation over  time slices was 0.5, 1.0, 0.2 for 
$\Gamma=\gamma_5$, $\gamma_5 \gamma_4$, I respectively for local
operators and the ratio was  about 50\% bigger for fuzzed hadronic 
operators).
  The computational effort  in terms  of the number of inversions is
equivalent to that in obtaining  two conventional propagators (from two
sources of  all colour-spins). We also make  use  of the
fact~\cite{sesam} that one can truncate the inversion at a larger
residual  without measurable bias, using $10^{-9}$ in an MR inverter as
residual/source. 

In conclusion,  using $N_S=24$ samples for each  gauge configuration
corresponds to 2 conventional propagator determinations (from all 12
colour-spin combinations) and so is not  a particularly big
computational challenge and yet the resulting measurement of the
disconnected  fermion loop has a stochastic error  which is unbiased and
less than its intrinsic error.

For spatial size 16, we made use of  existing solver codes and chose not
to use the variance reduction method described  above, using instead
$n_1=n_2=0$. To achieve  some variance reduction, we used the
method~\cite{fm} of using pairs of sources with the same random Z2
numbers  but with the second of the pair multiplied by $\gamma_i
\gamma_5$ where $i$ is chosen randomly from $1 \dots 3$. This has the
effect of reducing the standard error  by a factor of two for
$\Gamma=\gamma_5$ for only a doubling of CPU effort. In this case we
used either $N_S=48$ or 200, the larger value  being motivated by the
need to get a more precise estimate for  $\Gamma=\gamma_5 \gamma_4$.

\end{document}